\documentclass[11pt]{article}
\usepackage[scale={.75,.8}]{geometry}
\usepackage{amsmath}
\usepackage{amssymb}
\usepackage{graphicx}
\usepackage{subfigure}
\usepackage{cite}
\usepackage{slashed}
\usepackage{multirow}
\usepackage{color}

\newcommand{\beq}{\begin{equation}}
\newcommand{\eeq}{\end{equation}}

\newcommand{\ma}{\mathcal{A}}
\newcommand{\mc}{\chi}

\begin{document}
\title{
\begin{flushright}
{\small OUTP-1013P} \\
{\small CERN-PH-TH/2010-127}
\end{flushright}
\vspace{4mm}
{\bf Reconstructing events with missing transverse momentum at the LHC and its application to spin measurement}}
\author{Dean Horton\thanks{email: d.horton1@physics.ox.ac.uk} \vspace{4mm}\\
{\small \it The Rudolf Peierls Centre for Theoretical Physics, University of Oxford,}\\
{\small \it 1 Keble Road, Oxford, OX1 3NP, UK}\vspace{1mm}\\
{\small \it Department of Physics, CERN Theory Division,} \\
{\small \it CH-1211, Geneva 23, Switzerland }
}
\date{}
\maketitle

\abstract 
In this article we discuss the measurement of spin at the
LHC, in events with two unknown four-momenta. Central to this problem is
the identification of spin-dependent kinematic variables and the
construction of a statistical test that can distinguish between
different spin hypotheses. We propose a method for reconstructing
kinematic variables that depend upon the unknown momenta.
The method is based upon a probabilistic
reconstruction of each event, given the masses of the final and
intermediate states and the cross-section of the assumed
hypothesis. We demonstrate that this method can distinguish between
two spin hypotheses for a specific process, even after mass
uncertainties and Standard Model backgrounds are taken into
account. We compare our method with another that only utilises the
observable momenta of each event. We will show that our method permits an improved
discrimination between hypotheses, with a reduced probability of error.

\section{Introduction}
Since the end of March 2010 high-energy collisions have been taking
place at the Large Hadron Collider (LHC), ushering in a new era of
particle physics. Over the coming decades the experiments at the LHC
will test the Standard Model (SM) in a new energy regime and will
search for the first conclusive evidence for physics beyond it.
One possible signal for this new physics, which is motivated by the
existence of dark matter, are events with large missing transverse
momentum. Such events may have the following topology\footnote{In general, the event may also involve initial state radiation. We will assume that this can be neglected for the remainder of this paper.}:
\beq \label{e.Process} 
f\bar{f} \rightarrow \mathcal{A} + \bar{\mathcal{A}} \rightarrow
\mathcal{V}+\chi + \bar{\mathcal{V}}+\chi 
\eeq 
where constituents, $f\bar{f}$, of the colliding protons interact to
produce a pair of new particle states $\ma\bar{\ma}$. These
subsequently decay into an electrically neutral and colourless state
$\mc$ (the DM candidate), that will escape detection, and a set of
visible SM states $\mathcal{V}\bar{\mathcal{V}}$. Because the
four-momenta of the two $\mc$ particles are unknown, the kinematics of
these events cannot be reconstructed. This loss of information will
make it difficult to measure the masses, spins and couplings of these
particle states. Such measurements, however, will be crucial if the
physics underlying these signals is to be understood. In this paper we
focus on the issue of spin measurement in events of this topology.

In order to determine the spin of either $\ma$ or $\chi$, one must
measure the probability density of some kinematic quantity. This
distribution will then be compared to those predicted by several
hypotheses, e.g. for different spin assignments to the particles in
the event, and one must construct some statistical test in order to
decide which hypothesis is correct. Central to this problem,
therefore, is the identification of kinematic quantities, and an
associated statistical test, that will permit a spin determination
with a small probability of error. The ultimate goal is to identify an
`optimal' set, which will give the smallest probability of error for a
fixed luminosity.

Previous studies of spin determination at the LHC have only considered
kinematic quantities that are observable---i.e. those that can be
uniquely determined for each event. Examples include Lorentz
invariants formed from the visible momenta of decay
products~\cite{Athanasiou:2006ef,Barr:2004ze,Wang:2006hk,Burns:2008cp,Kramer:2009kp}, and
differences in pseudo-rapidities~\cite{Barr:2005dz}. These variables,
however, may not be optimal.

Alternative variables that depend upon the spin are the polar and azimuthal angles of
production and decay, measured relative to some physical axis in the
centre of mass frame and the rest frames of decaying particles. The
energy dependence of the cross-section is another spin-dependent
variable. It has been shown that using these variables one can
determine the spin uniquely, in certain
processes~\cite{Boudjema:2009fz, Choi:2006mr, Buckley:2008eb}. The
difficulty with these variables is that they depend upon the unknown
final state momenta, and so cannot be reconstructed event-by-event.
This does not imply, however, that one cannot obtain some information
regarding the probability density of these variables.

Consider some variable, $O$, with probability density $p(O|H)$
for some given hypothesis, $H$. (In principle $H$ must specify all
relevant physical quantities, including spins, masses and couplings.)
It follows that:
\beq\label{e.pOH}
p(O|H)= \sum_{V}p(O|V,H)p(V|H)
\eeq
where the sum is over each point, $V$, in the phase space of the
{\it visible} final state. Therefore, suppose we construct, for each
hypothesis, the following distribution for a given set of $N$ events:
\beq\label{e.qHdef}
q_H(O) \equiv \sum_{i=1}^{N}p(O|E_i, H)
\eeq
where the sum is over each of the events $E_i$ in the set. The
conditional probabilities $p(O|E_i,H)$ include all of the information
that we have about an event, including the observed visible momenta
and any information regarding the masses of $\ma$ and $\chi$. If $H$
is the true hypothesis, this distribution will have the following form
in the large statistics limit:
\beq\label{e.lim1}
\lim_{N \rightarrow \infty} q_H(O) \rightarrow \sum_{V}p(O|V,H)\langle N_V\rangle
\eeq
where the sum over events has been rearranged into a sum over phase space points, with $\langle N_V \rangle$ the average number of events at the point $V$. Since $\langle N_V \rangle = N p(V|H)$, it follows from Eqns.~\eqref{e.pOH} and \eqref{e.lim1} that
\beq
\lim_{N \rightarrow \infty} q_H(O) \rightarrow N p(O|H)
\eeq
In general, $q_H$ will tend to a different distribution if $H$ is a
false hypothesis.  Hence, in principal, one can identify the true
hypothesis by comparing the distribution $q_H(O)$ with $Np(O|H)$. It
is this idea that we will explore in this paper.

The remainder of this paper is organised as follows. In
Sec.~\ref{s.Method} we will demonstrate how one can construct, for a
given hypothesis and kinematic variable, the distribution $q_H(O)$. We
will then discuss how to construct a statistical test to compare
different hypotheses and decide which is true.

In Sec.~\ref{s.Test} we will apply our method to the discrimination of
SUSY and UED models in slepton pair-production. For reference, we will
compare our method with another that currently exists in the
literature.  We note that the purpose of this study is not to perform
a fully rigorous analysis of how accurately our method will perform at
the LHC. Such a study will require a full understanding of the
detector and backgrounds, combined with a detailed study of systematic
uncertainties. This is beyond the scope of this work. Rather, our goal
is to perform a preliminary study, in a somewhat idealised case, in
order to potentially identify the best method to use for a spin
measurement.

Finally, in Sec.~\ref{s.Conclusion} we will summarise our results and
make some suggestions for further developments.

\section{The method} \label{s.Method}

This section is composed of two parts. In the first we will discuss
how to construct the distribution $q_H$ for a set of events and a
given hypothesis $H$. Then, in the second part, we will discuss how
$q_H$ can be used in a statistical test to discriminate between
different hypotheses.

\subsection{Constructing $q_H$}
\begin{figure}[t]
  \centering
  \begin{picture}(200,100)
    \put(3, 50){$p_A$}
    \put(20,53){\vector(1,0){50}}
    \put(150,53){\vector(-1,0){50}}
    \put(155, 50){$p_B$}
    \put(85,53){\circle*{5}}
    \put(85,53){\vector(1,1){15}}
    \put(85,53){\vector(-1,-1){15}}
    \put(100,68){\vector(0,1){30}}
    \put(100,68){\vector(2,1){20.8}}
    \put(70,38){\vector(0,-1){30}}
    \put(70,38){\vector(-2,-1){20.8}}
    \put(80, 64){$\ma$}
    \put(85, 90){$\mathcal{V}$}
    \put(110, 64){$\chi$}
    \put(78,35){$\bar{\ma}$}
    \put(73, 15){$\bar{\mathcal{V}}$}
    \put(48,39){$\chi$}
    \put(98, 105){$p$}
    \put(125, 78){$k$}
    \put(40,18){$\bar{k}$}
    \put(67, 0){$\bar{p}$}
  \end{picture}
  \caption{\small Defines the convention for momentum assignment in the process of Eq.~\eqref{e.Process}.}
  \label{f.MomLabels}
\end{figure}
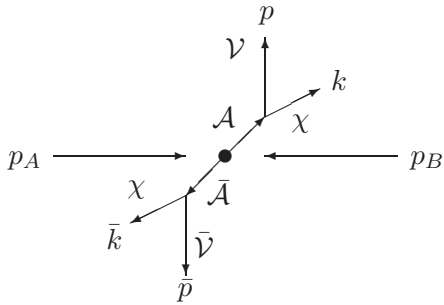
Let us consider a set of $N$ events with the topology of Eq.~\eqref{e.Process}, where the initial and final state momenta are labelled as shown in Fig.~\ref{f.MomLabels}. From this set we wish to construct the distribution $q_H(O)$, defined in Eq.~\eqref{e.qHdef}, for some kinematic quantity $O$ and hypothesis $H$. To do this we perform the following procedure. For each event, $E_i$:
\begin{itemize}
\item We determine the space of all possible four-momenta $k$ and $\bar{k}$ that can be assigned to  $E_i$, consistent with the hypothesis $H$. We will call this the space of solutions.
\item We construct the probability density $p(k,\bar{k}|E_i,H)$ on this space.
\item We integrate $p(k,\bar{k}|E_i,H)$ over the space of solutions to determine the distribution $p(O|E_i,H)$.
\end{itemize}
We then sum the $p(O|E_i,H)$ from each event to obtain $q_H$. We will now discuss each of these points in more detail.

\subsubsection{The space of solutions}\label{s.Solutions}
The final state of an event $E_i$ is parameterised by the eight unknown components of the two four-vectors $k$ and $\bar{k}$. Not all of these components are independent. Due to the conservation of four-momentum there exists the constraint:
\beq \label{e.C4mom}
p_A+p_B=(E, 0, 0, P_{\rm z}) = p + k + \bar{p} + \bar{k}
\eeq
where $E$ and $P_{\rm z}$ are the (unknown) total energy and momentum, respectively, in the LAB frame. Hence the transverse components of $k$ and $\bar{k}$ must satisfy:
\beq\label{e.Ctransverse}
\vec{k}_T+\vec{\bar{k}}_T=\slashed{\vec{p}}_T\equiv-(\vec{p}_T+\vec{\bar{p}}_T)
\eeq

We will further assume that the masses $m_{\mathcal{A}}$ and $m_{\chi}$, of the states $\mathcal{A}$ and $\chi$ respectively, have been measured. 
Several methods have been proposed to measure these masses in a model independent manner, and we refer the reader to the relevant literature~\cite{Barr:2010zj}. 
Hence, knowing the mass, one can apply the on-shell constraint to the unknown momenta:
\begin{align}  \label{e.CmChi}
  k^2&=m_{\chi}^2 & \bar{k}^2&=m_{\chi}^2  
\end{align}

The intermediate state $\mathcal{A}$ will have some finite width, $\Gamma_{\ma}$.
It is unlikely that a reliable measure of $\Gamma_{\ma}$ will be possible at the LHC. Given this we will make the narrow width approximation for this state, which has the additional advantage of providing two further on-shell constraints:
\begin{align}  \label{e.CmA}
  (p+k)^2&=m_{\mathcal{A}}^2 & (\bar{p}+\bar{k})^2&=m_{\mathcal{A}}^2 
\end{align}
Provided $\Gamma_{\ma}/m_{\ma} \ll 1$, this is likely to be a good approximation.

The six constraints given by Eqns.~\eqref{e.Ctransverse}, \eqref{e.CmChi} and \eqref{e.CmA} reduce the number of free parameters in the final state to two. Without loss of generality we choose these free parameters to be the transverse components $\vec{k}_T$. By solving the constraint equations one finds, for a given $\vec{k}_T$, two solutions for the four-momenta $k$:
\begin{align}
k(i)&=(k^0(i), \vec{k}_T, k_L(i) ) \\
k^0(i)&=\sqrt{m_\chi^2+ \vec{k}_T\cdot \vec{k}_T + k_L^2(i)} \\
k_L(i)&=p_L\bigg(\frac{k^0_T}{p^0_T}+\frac{\delta m^2}{2(p^0_T)^2}\bigg)+(-1)^i p^0 \frac{\sqrt{\delta m^2}\sqrt{2p_T^0k_T^0+\delta m^2}}{2(p^0_T)^2} \label{e.SolnskL}
\end{align}
where $i\in\{0,1\}$. We have defined $\delta m^2=m_{\mathcal{A}}^2-m_T^2(p,k)$, where $m_T(a,b)$ is the transverse mass of the four-vectors $a$ and $b$:
\beq
m_T^2(a,b)=a^2+b^2+2(a^0_Tb^0_T-\vec{a}_T\cdot\vec{b}_T)
\eeq
and $a_T^0=\sqrt{a^2+\vec{a}_T\cdot \vec{a}_T}$ is the transverse energy.
Through the substitutions $p\rightarrow \bar{p}$ and $k\rightarrow \bar{k}$, with $\vec{\bar{k}}_T=\slashed{\vec{p}}_T-\vec{k}_T$, one obtains the equivalent solutions for $\bar{k}$. For each choice of $\vec{k}_T$ there are four discrete solutions, corresponding to the two different sign choices for $k_L$ and $\bar{k}_L$.

A real solution exists only if the two following constraints are satisfied: $m_T^2(p,k)\leq m_{\mathcal{A}}^2$ and $m_T^2(\bar{p},\bar{k})\leq m_{\mathcal{A}}^2$. By choosing an orthogonal set of co-ordinate axes in the transverse plane such that $\vec{p}_T=|\vec{p}_T|(1,0)$, the first constraint can be written in the form:
\beq\label{e.C1}
p^2k_x^2+(p_T^0)^2k_y^2-2 c_1|\vec{p}_T|k_x - c_0 \leq 0
\eeq
where the constants $c_0$ and $c_1$ are given by:
\begin{align}
c_1 &= \tfrac{1}{2}(m_{\mathcal{A}}^2 - m_{\chi}^2 - p^2) \\
c_0 &= c_1^2 - (p_T^0)^2m_{\chi}^2
\end{align}
whilst the second has the form:
\begin{align}\label{e.C2}
(\bar{p}_T^0)^2\vec{k}_T\cdot\vec{k}_T-(\vec{\bar{p}}_T\cdot\vec{k}_T)^2
+2\vec{d}_1\cdot \vec{k}_T + d_0 \leq 0
\end{align}
where
\begin{align}
\vec{d}_1 &= (\bar{c}_1+\vec{\bar{p}}_T\cdot\slashed{\vec{p}}_T)\vec{\bar{p}}_T - (\bar{p}_T^0)^2\slashed{\vec{p}}_T\\
d_0 &= (\bar{p}_T^0)^2\slashed{\vec{p}}_T\cdot\slashed{\vec{p}}_T-(\vec{\bar{p}}_T\cdot\slashed{\vec{p}}_T)^2-2\bar{c}_1\vec{\bar{p}}_T\cdot \slashed{\vec{p}}_T-\bar{c}_0
\end{align}
Thus $\vec{k}_T$ is constrained to lie within the overlap region of two ellipses, which are defined by Eqns.~\eqref{e.C1} and \eqref{e.C2}. Thus the space of solutions associated with an event $E_i$ is the set of four discrete solutions associated with each $\vec{k}_T$ that lies within this overlap region. 

\subsubsection{Constructing a probability measure on the space of solutions}
The probability measure $p(k,\bar{k}|E_i, H)$ is given by the differential cross-section:
\beq
d\sigma = \int_0^1 dx \int_0^1 dx' \sum_{a,b} f_a(x) f_{b}(x') d \hat{\sigma }_{ab}
\eeq
where $f_a$ are the proton's parton density functions (PDFs) and the sum is performed over all parton flavours $a$. We define $x$ and $x'$ such that:
\begin{align}
p_A &= E_b x (1, 0,0,1) \\
p_B &= E_b x' (1,0,0,-1)
\end{align}
where $E_b$ is the beam energy. For the LHC we will assume the full design beam energy of 7 TeV.
\newline
\newline
The cross-section $d\hat{\sigma}$ for the hard scatter is given by:
\beq\label{e.HSxsec}
d\hat{\sigma}_{ab}=\frac{1}{2s}\bigg(\prod_{i=1}^{N}\frac{d^3\vec{a}_i}{(2\pi)^32a_i^0}\bigg)(2\pi)^4\delta^4(p_A+p_B-\sum_ia_i)|\mathcal{M}_{ab}|^2
\eeq
where $\sqrt{s}$ is the total energy in the CM frame and a general four-momentum in the final state is written $a_i$. The matrix element for the process, which is specified for a given hypothesis $H$, is $\mathcal{M}_{ab}$.
\newline
\newline
Making the narrow width approximation for the intermediate state $\mathcal{A}$ forces $\mathcal{M}_{ab}$ to have the following structure:
\beq\label{e.NWA}
|\mathcal{M}_{ab}|^2 = |\hat{\mathcal{M}}_{ab}|^2\delta \bigl( (p+k)^2-m^2_{\mathcal{A}} \bigr) \delta \bigl( (\bar{p}+\bar{k})^2 - m^2_{\mathcal{A}} \bigr)
\eeq
where $\hat{\mathcal{M}}_{ab}$ is a dimensionless quantity.
Inserting Eq.~\eqref{e.NWA} into Eq.~\eqref{e.HSxsec} and integrating  over the components of $\vec{\bar{k}}$ and $k_L$, one finds, up to irrelevant constant factors:
\beq
d\hat{\sigma}_{ab}\propto\frac{d^3\vec{p}}{p^0p^0_T}\frac{d^3\vec{\bar{p}}}{\bar{p}^0\bar{p}^0_T}\sum_{k_L, \bar{k}_L}\frac{|\hat{\mathcal{M}}_{ab}|^2}{s}\cdot\frac{d^2\vec{k}_T}{k_T^0\bar{k}_T^0}\frac{1}{|\sinh(\Delta y)\sinh(\Delta\bar{y})|}\delta^{(2)}(p_A+p_B-\sum_i p_i)
\eeq
where the sum over $k_L$ and $\bar{k}_L$ corresponds to a sum over the solutions given in Eq.~\eqref{e.SolnskL}. The rapidity differences $\Delta y$ and $\Delta \bar{y}$ are given by:
\begin{align}
\Delta y &= \frac{1}{2}\ln \bigg(\frac{p^0 + p_L}{p^0 - p_L}\cdot\frac{k^0 - k_L}{k^0 + k_L}\bigg) &
\Delta \bar{y} &= \frac{1}{2}\ln \bigg(\frac{\bar{p}^0 + \bar{p}_L}{\bar{p}^0 - \bar{p}_L}\cdot\frac{\bar{k}^0 - \bar{k}_L}{\bar{k}^0 + \bar{k}_L}\bigg)
\end{align}
Hence the total differential cross-section has the form (again up to constant factors):
\beq
d\sigma \propto \sum_{k_L, \bar{k}_L}\frac{\sum_{a,b} f_a(x)f_{b}(x')|\hat{\mathcal{M}}_{ab}|^2}{s}\cdot\frac{d^2\vec{k}_T}{k_T^0\bar{k}_T^0}\frac{1}{|\sinh(\Delta y)\sinh(\Delta\bar{y})|}
\eeq
where $x$ and $x'$ are fixed by the requirement:
\begin{gather}
E_b(x+x',0,0,x-x') = p+k+\bar{p}+\bar{k} \\
0\leq x,x'\leq 1
\end{gather}
Hence the probability measure $p(\vec{k}_T, i,j|E,H)$ is given by:
\beq\label{e.ProbMeasure}
p(\vec{k}_T,i,j|E,H)d^2\vec{k}_T = \mathcal{N}\frac{\sum_{a,b}f_a(x)f_{b}(x')|\hat{\mathcal{M}}_{ab}|^2}{xx'}\cdot\frac{d^2\vec{k}_T}{k_T^0\bar{k}_T^0|\sinh(\Delta y)\sinh(\Delta\bar{y})|}
\eeq
where $i,j\in\{0,1\}$ correspond to the different $k_L$ and $\bar{k}_L$ solutions at each point. The constant $\mathcal{N}$ is defined such that $\sum_{i,j}\int p(\vec{k}_T,i,j|E,H)d^2\vec{k}_T=1$. 

\subsubsection{Constructing $p(O|E,H)$}
Given the space of solutions defined in Sec.~\ref{s.Solutions} and the probability measure of Eq.~\eqref{e.ProbMeasure}, one can obtain the probability density for any kinematic quantity $O$ through a straightforward change of variables. In practice we simply integrate $p(k,\bar{k}|E,H)$ over the region of solution space for which $O$ takes a given value:
\beq
p(\mathcal{O}'|E,H)=\int d\vec{k}_T\sum_{i,j}p(\vec{k}_T,i,j|E,H)\delta(\mathcal{O}(\vec{k}_T)-\mathcal{O}')
\eeq
where the integral and sum is performed over the entire space of solutions.

\subsection{The statistical test}\label{s.StatTest}
The previous sections have detailed how to construct $q_H$ for a set
of $N$ events. Given this distribution for each hypothesis, one would
then like to decide which hypothesis is true.
For this, we must construct a statistical test.

Since $q_H$ depends upon the random events in the set, $q_H$ is itself
random and so will be distributed according to some probability
density. For events distributed according to the hypothesis $H$ let us
denote the probability density of $q_H$ as $f(q_H|H)$. (That is,
$f(q_H|H)$ is the probability density for $q_H$ in events generated
and reconstructed with the same hypothesis. In statistics terminology, one can also consider $f(q_H|H)$ to be a likelihood.) Let us assume that there
are $l$ hypotheses $H_i$, with $i\in\{1, \dotsb, l\}$, each with an
associated probability density $f(q_{H_i}|H_i)$. If we knew each of
these probability densities we could construct a statistical test as
follows.

Suppose that we are given a set of $N$ events, and we are asked to
determine which of the $l$ hypotheses is true. For each hypothesis we
could calculate $q_{H_i}$, as detailed above, and then calculate the
associated probability, given the hypothesis, $f_i =
f(q_{H_i}|H_i)$. We then reject the hypothesis $H_j$ in
favour of $H_i$ if
\beq
r_{ij} \equiv \frac{f_i}{f_j} > c_{ij}
\eeq
where the constants $c_{ij}$ are chosen so as to minimise errors (i.e. rejecting $H_i$ when $H_i$ is true, or accepting $H_i$ when $H_i$ is false). We will call $r_{ij}$ the likelihood ratio.

Unfortunately we cannot calculate the probability densities $f(q_H|H)$
analytically. Instead, they will be determined numerically, through
Monte Carlo simulation.  To simplify matters we will divide $O$ into
$M$ discrete bins, where $M$ is made sufficiently small so as to
simplify the calculation, without losing significant detail
in the shape of $q_H$. Thus for each set of events one
calculates the $M$ discrete bin values $q_H^i$, with $i\in\{1,\dotsb,
M\}$. We will denote this set of $M$ values by the vector
$\vec{q}_H\equiv\{q_H^1, \dotsb, q_H^M\}$. Hence the problem of
determining $f$ is reduced to finding the $M$-dimensional probability
distribution $f(\vec{q}_H|H)$.

Since $q_H^i$ is the sum over many random variables:
\beq
q_H^i = \sum_{j=1}^N p(O^i|E_j,H)
\eeq
one expects from the central limit theorem that it will have a Gaussian distribution. (This will hold provided $N \gg 1$. For any process that may be viably studied at the LHC one might expect $N\sim \mathcal{O}(10^2)$, so this assumption seems reasonable.) It seems sensible therefore to assume an $M$-dimensional Gaussian distribution for $f(\vec{q}_H|H)$:
\beq
f(\vec{q}_H|H) = \sqrt{\frac{\det V_H^{-1}}{(2\pi)^M}} \exp\Big(-\tfrac{1}{2}(\vec{q}_H-\vec{\mu}_H)^{T}V_H^{-1}(\vec{q}_H-\vec{\mu}_H)\Big)
\eeq
where $\mu_H^i$ is the average value in each bin and $V_H$ is the $M\times M$ correlation matrix. In general, since $q_H$ is a smooth distribution, there will be significant correlations between the bin values. The $\tfrac{1}{2}M(3+M)$ parameters of this Gaussian distribution will be estimated using Monte Carlo simulation. We note, however, that it follows from the definition of $q_H$ that:
\beq
 \mu^i = \langle q_H^i \rangle = \langle N\rangle p(O^i|H)
\eeq
which provides an important cross-check of our method.

\subsection{Summary}
In this section we have discussed how one can construct $q_H(O)$ and
then use this in a statistical test to discriminate between
hypotheses.  In order to illustrate the method we apply it to an
example in the following section.

\section{A test of the method: Discriminating between models with different spin}\label{s.Test}
\begin{table}[t]
  \begin{center}
  \begin{tabular}{|c|c|c|c|c|c|c|}
    \hline 
    Model & $\ma$          & $S_{\ma}$ &  $R(\ma)$       & $\mc$            & $S_{\mc}$  &  $R(\mc)$             \\
    \hline                                                                                                 
    SUSY & $\tilde{\mu}_L$ & $0$       &  $(0,2,1/2)$    & $\chi_1^0$       & $1/2$      &  $(0,0,0)$           \\
    UED  & $\mu_L^{(1)}$   & $1/2$     &  $(0,2,1/2)$    & $B_{\mu}^{(1)}$  & $1$        &  $(0,0,0)$           \\ 
    \hline
  \end{tabular}
  \end{center}
  \caption{\small Table of spin $S$ and representation $R$ assignments in the SUSY and UED models. (Note that in UED $\ma$ is a Dirac field, with both left and right handed components transforming in the same representation.)}
  \label{t.ParticleAssignment}
\end{table}

As a first test of the method we have attempted to discriminate
between two models in the following process:
\beq\label{e.TestProcess}
q \bar{q} \rightarrow \ma \bar{\ma} \rightarrow \mu^- + \chi + \mu^+ + \chi
\eeq
The models differ in their spin assignments and are based upon:
supersymmetry (SUSY), where $\ma$ is the scalar superpartner of the
muon and $\chi$ is a Majorana fermion, and universal extra dimensions
(UED), where $\ma$ and $\chi$ are the first Kaluza-Klein excitation of
the muon and $B$ gauge boson, respectively. The spin $S$ and
representation $R$ under the SM gauge group $SU(3)\times SU(2) \times
U(1)_Y$ of $\ma$ and $\mc$ are summarised for each model in
Table~\ref{t.ParticleAssignment}.

Several previous works have shown that a (partial) measurement of the
spin of $\ma$ is feasible in this process, both at a linear collider
and at the LHC~\cite{Choi:2006mr, Barr:2005dz, Buckley:2008eb}. Let us
summarise, briefly, how this can be achieved.

The $q\bar{q}$ initial state of this process is forced to have a total
angular momentum $J_z=\pm 1$. (This is a consequence of the antiparallel
momenta of the $q$ and $\bar{q}$ together with their coupling to an
s-channel vector boson.) Hence, in the SUSY model, the scalar pair
$(\ma\bar{\ma})$ must be produced in a $P$-wave state, in order to
conserve angular momentum.  These states carry orbital angular
momentum and have a non-trivial angular distribution. Thus the SUSY
matrix element has the form:
\beq\label{e.meSUSY}
|\hat{\mathcal{M}}|^2_{\rm SUSY} = \kappa_{\rm SUSY} \beta_{\ma}^2 \sin^2\theta_{\ma}
\eeq
where $\theta_{\ma}$ and $\beta_{\ma}$ are the polar angle of the
$\ma$ momentum, measured relative to the beam-line, and the speed of
$\ma$, as observed in the CM frame. The factor $\kappa$ is
dimensionless and independent of $\theta_{\ma}$. This matrix element
vanishes in the limits $\sin\theta_{\ma}\rightarrow 0$ and
$\beta_{\ma} \rightarrow 0$ as a straightforward consequence of
angular momentum conservation. If, however, $\ma$ carries intrinsic
angular momentum through a non-zero spin, total angular momentum can
be conserved even with vanishing orbital angular momentum. Hence the
$(\ma\bar{\ma})$ can be produced in an $S$-wave state, which has a
matrix element that is non-zero for all $\theta_{\ma}$ and
$\beta_{\ma}$. The precise form of the matrix element depends upon the
spin and couplings of $\ma$. If $\ma$ is spin-1/2 with vector-like
couplings, as in UED, this matrix element has the form:
\beq \label{e.meUED}
|\hat{\mathcal{M}}|^2_{\rm UED} = \kappa_{\rm UED} (2-\beta_{\ma}^2 \sin^2\theta_{\ma})
\eeq
Thus the distributions of both $\cos\theta_{\ma}$ and $\beta_{\ma}$
are sensitive to the spin and can be used to discriminate between the
SUSY and UED models. A more general analysis has shown that for a
process of this kind, these distributions are sufficient to
discriminate between the hypotheses $S_{\ma}=0$ and $S_{\ma}>0$
\cite{Choi:2006mr}.

\begin{figure}[t]
  \centering
  \subfigure[SUSY]{
    \includegraphics[width=0.45\textwidth]{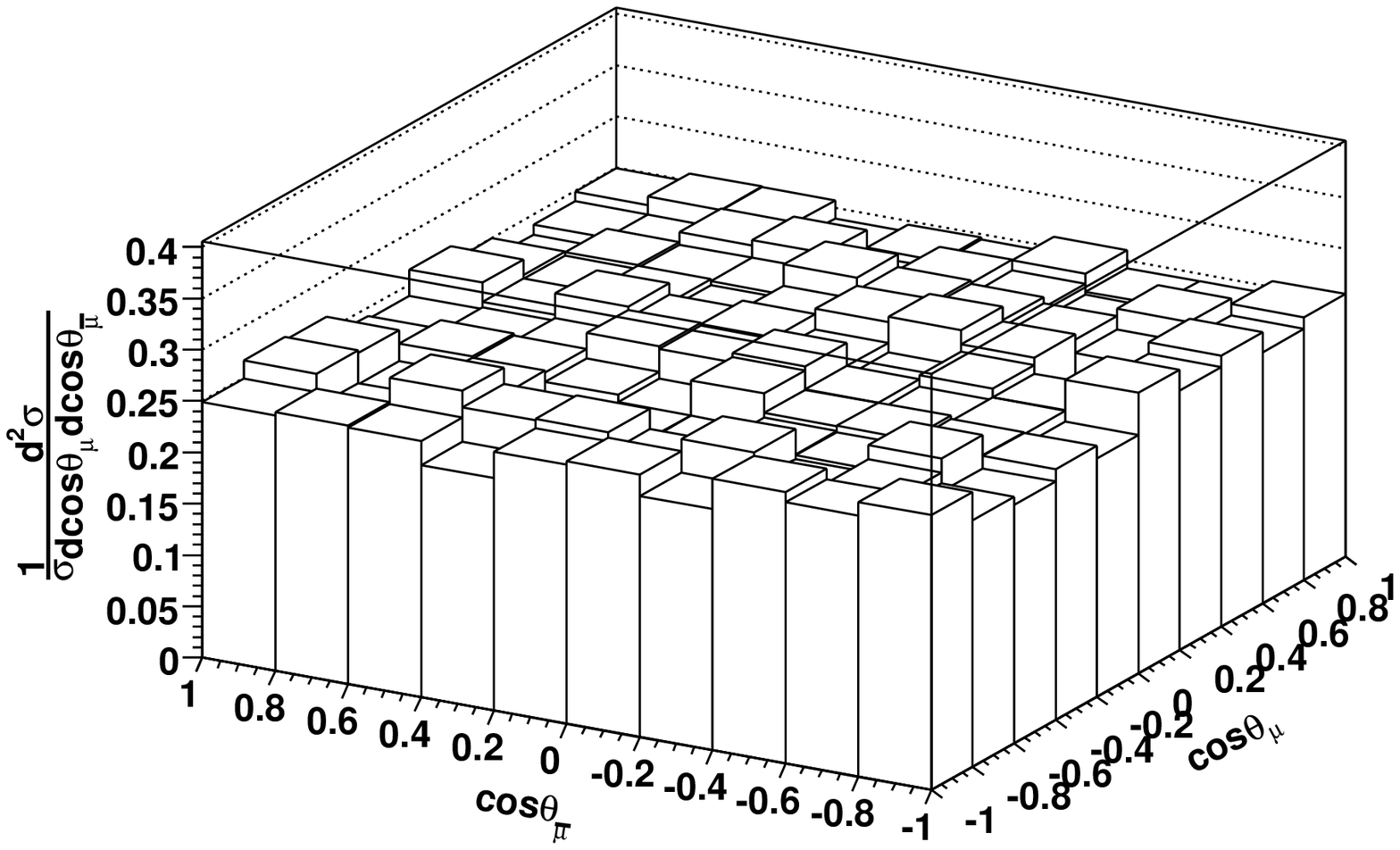}
    \label{f.thetaMu.susy}
  }
  \subfigure[UED]{
    \includegraphics[width=0.45\textwidth]{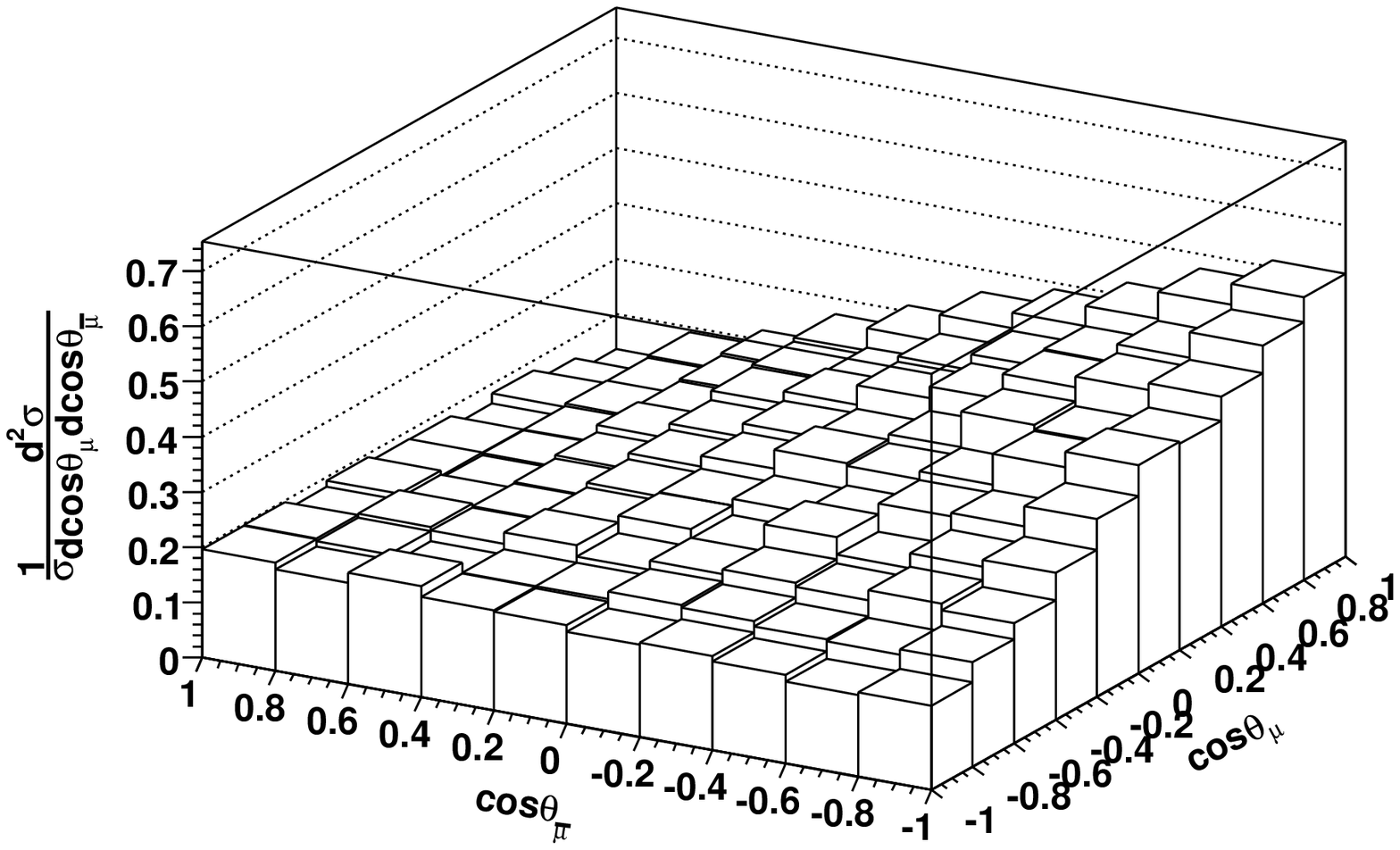}
    \label{f.thetaMu.ued}
  }
  \caption{\small A plot of the $\cos\theta_{\mu}$---$\cos\bar{\theta}_{\mu}$ distribution in the SUSY and UED models.}
  \label{f.thetaMu}
\end{figure}
Other variables are also sensitive to the spin, such as the azimuthal and polar angles of the $\mu^-$ ($\mu^+$) momenta in the rest frame of the $\ma$ ($\bar{\ma}$). Of particular interest are the following:
\begin{itemize}
\item The azimuthal angle $\phi$ between the decay planes of the $\ma$ and $\bar{\ma}$, defined in the CM frame. As pointed out in~\cite{Buckley:2008eb}, $\phi$ has a distribution of the form:
\beq
\frac{1}{\sigma}\frac{d\sigma}{d\phi} = \frac{1}{2\pi}\sum_{i=0}^{2S_{\ma}}(A_i\cos2i\phi)
\eeq
Hence one can infer the spin of $\ma$ from the highest $\cos(2i\phi)$ component of the distribution.
\item The polar angles $\theta_{\mu}$ ($\bar{\theta}_{\mu}$) of the $\mu^-$ ($\mu^+$), measured in the rest frame of the $\ma$ ($\bar{\ma}$). These are measured relative to the direction of the boost from the LAB to the CM frame, as viewed in the appropriate rest frame. In the rest frame of $\ma$ only its polarisation can break rotational invariance. Hence the observation of a non-trivial distribution in $\theta_{\mu}$ and $\bar{\theta}_{\mu}$ indicates that $\ma$ must be polarised and hence have a non-zero spin. For the UED model there is a large correlation\footnote{This correlation arises because the decay vertex involves a chiral coupling and there is a net polarisation of the initial state along the direction of the boost, $\vec{B}$, from the CM frame to the LAB frame. This polarisation arises as follows. Because $\ma$ is a doublet under $SU(2)$ the process $q\bar{q}\rightarrow \ma\bar{\ma}$ involves a significant contribution from s-channel $Z$ exchange. Since the $Z$ has chiral couplings to the quarks, this results in an overall polarisation of the initial state along the direction defined by the incoming quark's momentum. However, owing to the structure of the pdfs and the mass of $\ma$, this direction is correlated with the direction of the boost $\vec{B}$.} between these angles, whilst the SUSY distribution is flat, as shown in Fig.~\ref{f.thetaMu}.
\end{itemize}

Unfortunately, all of these variables depend upon the unknown $\chi$
momenta, so they cannot be uniquely determined for an event at the
LHC. However, we propose to employ our method to reconstruct their
distributions and so discriminate between models.

An alternative method has been proposed for use at the LHC, which depends upon the variable $\cos\theta_{ll}$~\cite{Barr:2005dz}:
\beq
\cos\theta_{ll} = \tanh\bigg(\frac{\Delta\eta_{+-}}{2}\bigg)
\eeq
where $\Delta\eta_{+-}$ is the pseudorapidity difference between the $\mu^+$ and $\mu^-$. This variable depends only upon the observable momenta and so is straightforward to reconstruct. Because it depends upon a pseudorapidity difference, it is invariant under longitudinal boosts and so is sensitive to the direction of the $\mu^{\pm}$ system in the CM frame. Since these states tend to be boosted in the direction of their parent $\ma$  and $\bar{\ma}$ this variable is correlated with the polar angle $\theta_{\ma}$, and so is sensitive to the spin. We propose to use this method as a benchmark for our own.

\subsection{Simulating events at the LHC - signal, background and event selection}
The LHC experiments will search for events from this process using the signature of two opposite sign muons, in association with missing transverse energy. There will, however, be significant background sources of such events arising from SM processes. 

A realistic test of our method must take into account the effect of these backgrounds. However, a complete background analysis, combined with detector effects and associated systematic errors is far beyond the scope of this work. Instead we will perform our test only at parton level, without hadronisation and detector simulation. Furthermore, we will only include a limited number of backgrounds in this analysis. Previous studies~\cite{Barr:2005dz} have identified several important sources, the most important of which is the irreducible SM background:
\beq\label{e.SMbkg}
q\bar{q} \rightarrow \mu^+ + \mu^- + \nu_{\mu} + \bar{\nu}_{\mu}
\eeq
This is the only one that we will take into account\footnote{Using the cuts listed below, it has been found~\cite{Barr:2005dz} that this process accounts for over $80\%$ of the background signal.}. Despite these simplifications this test should still establish whether, in principal, our method improves upon the use of the $\cos\theta_{ll}$ distribution.
 
In order to isolate the signal events from background we follow the work of Barr~\cite{Barr:2005dz} and apply the following selection criteria:

\begin{itemize}
\item The transverse momenta of the muons must satisfy $p_{T1} > 40\text{ GeV}$ and $p_{T2}>30\text{ GeV}$, where, by definition, $p_{T1}>p_{T2}$.
\item The invariant mass of the muon pair must satisfy: $m_{\mu\bar{\mu}} > 150 \text{ GeV}$ 
\item The missing transverse energy $\slashed{E}_T$ must satisfy $\slashed{E}_T > 100 \text{ GeV}$ 
\item $M_{T2}(M = 0) > 100 \text{ GeV}$
\end{itemize}
where the kinematic variable $M_{T2}$ \cite{Lester:1999tx,Barr:2003rg} is defined as:
\beq
M_{T2}(M)= \min_{\vec{k}_T}\bigg(\max \big(m_T(p,k), m_T(\bar{p},\bar{k})\big)\bigg)
\eeq
where the minimisation is subject to the constraints $k^2=\bar{k}^2=M^2$ and $k_T+\bar{k}_T=\slashed{p}_T$. 
These cuts have been developed to eliminate most of the events from the SM di-boson processes: $W^+W^-$ and $ZZ$ production, but have not been tuned to maximise the ratio of signal to background. 
In addition to these criteria we also demand that the final state muons have a pseudorapidity $|\eta| < 2.5$, which is required to ensure a high reconstruction efficiency in both ATLAS and CMS~\cite{Aad:2009wy, Acosta:922757}.

Using MadGraph/MadEvent~\cite{Alwall:2007st} we find that the leading order cross-section for the SM process given in Eq.~\eqref{e.SMbkg} is, after applying the above cuts,
\beq \label{e.xsecSM}
\sigma_{\rm SM}=0.5 \text{ fb}
\eeq

The signal events for both the SUSY and UED model were also generated using MadGraph/MadEvent. The UED model was implemented as a user-defined model, with particle widths calculated using the program BRIDGE \cite{Meade:2007js}. For this test we fixed the masses at $m_{\ma} = 350 \text{ GeV}$ and $m_{\chi} = 50 \text{ GeV}$.
For the SUSY model we found a leading order cross-section, after cuts, of 
\beq\label{e.susy_xsex}
\sigma_{\rm SUSY}=1.0 \text{ fb.} 
\eeq
whilst the cross-section for the UED model takes a significantly larger value. Because the couplings of the new particle states cannot be known prior to a spin determination, the cross-section cannot be used as true spin (or model) discriminant. (Our method only relies upon the shape of the $q_H$ distributions to perform the discrimination.) Thus we simulate both SUSY and UED models by assuming the same cross-section for each and, to be conservative, we choose the smaller cross-section of the SUSY model. 

\subsection{Implementing and testing our method}\label{s.Implementation}
For this test we performed a total of 2000 pseudo-experiments, each with a total integrated luminosity $\mathcal{L} = 100 \text{ fb}^{-1}$, for each model. In each pseudo-experiment the number of signal and background events were selected at random from Poisson distributions with means $\langle N_S \rangle = 100$ and $\langle N_B \rangle = 50$, respectively. The corresponding number of signal and background events were then generated and reconstructed using our method.

When calculating $q_H$ for each hypothesis we made several simplifications in the probability measure $p(k, \bar{k}|E,H)$, given in Eq.~\eqref{e.ProbMeasure}. Rather than use the full matrix element for either hypothesis, we instead used the matrix element for the production process only, which are given in Eqns.~\eqref{e.meSUSY} and \eqref{e.meUED}. The $\kappa$ factors were treated as dimensionless constants and were absorbed into the normalisation factor $\mathcal{N}$. 
In addition, we only include the PDFs for the up-type quarks\footnote{We have found that there is no significant difference between using the PDFs for either the up-type or down-type quarks.}, which were calculated using the NNLO central MSTW set~\cite{Martin:2009iq}, at a scale $Q=m_{\ma}$. 
These approximations were motivated by the desire to lessen the dependence of the reconstruction on unknown, mode-dependent factors, such as the structure of decay vertices and the strengths of couplings.
With these approximations we are effectively left with with a probability measure $p(k, \bar{k}|E,H)$ that is independent of these unknown factors. The approximations neglect some very mild dependence on $\sqrt{s}$ that is actually present in the $\kappa$ factors, and spin correlations in the decays of $\ma$ are also neglected. 

Because the number of signal and background events expected to pass the selection criteria are comparable, the reconstructed $q_H$ distribution will contain a significant background contamination. 
A background subtraction was performed in order to isolate the signal contribution to $q_H$. 
To do this we determined the average background contribution, $\langle \tilde{q}_H \rangle$, which was then subtracted in each pseudo-experiment to obtain a new distribution:
\beq
\hat{q}_H = q_H - \langle \tilde{q}_H\rangle
\eeq
These subtracted distributions were then used to perform the statistical tests discussed in Sec.~\ref{s.StatTest}. In particular, they were used to determine the means and the covariance matrix for the probability densities $f(\vec{\hat{q}}_{\rm SUSY}|{\rm SUSY})$ and $f(\vec{\hat{q}}_{\rm UED}|{\rm UED})$, using unbiased estimators.  
These probability densities were then used to calculate, for each pseudo-experiment, the likelihood ratio
\beq
r=\frac{f(\vec{\hat{q}}_{\rm SUSY}|{\rm SUSY})}{f(\vec{\hat{q}}_{\rm UED}|{ \rm UED})}
\eeq

\subsection{Results}
\begin{figure}[p]
  \centering
  \subfigure[$H=\text{SUSY}$]{
    \includegraphics[width=0.85\textwidth]{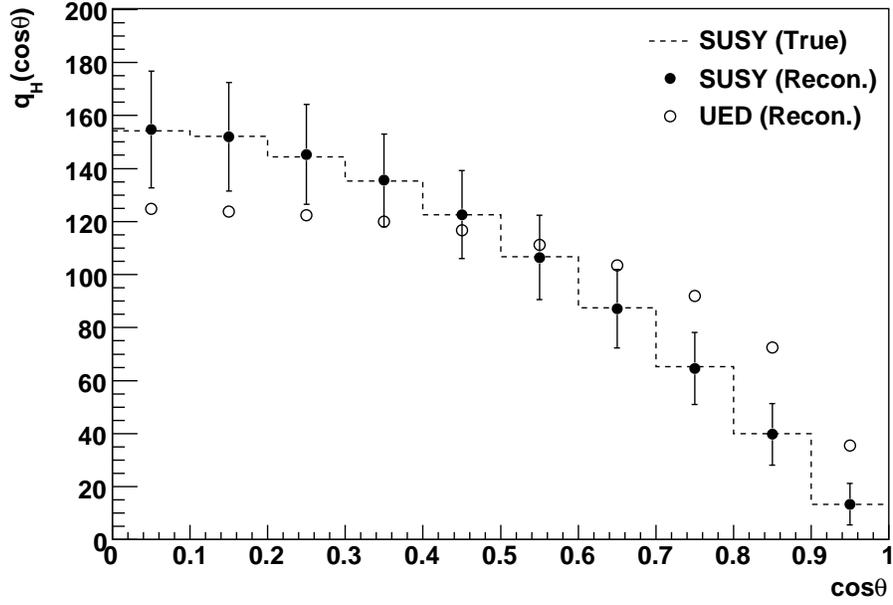}
    \label{f.cA.susy}
  }
  \subfigure[$H=\text{UED}$]{
    \includegraphics[width=0.85\textwidth]{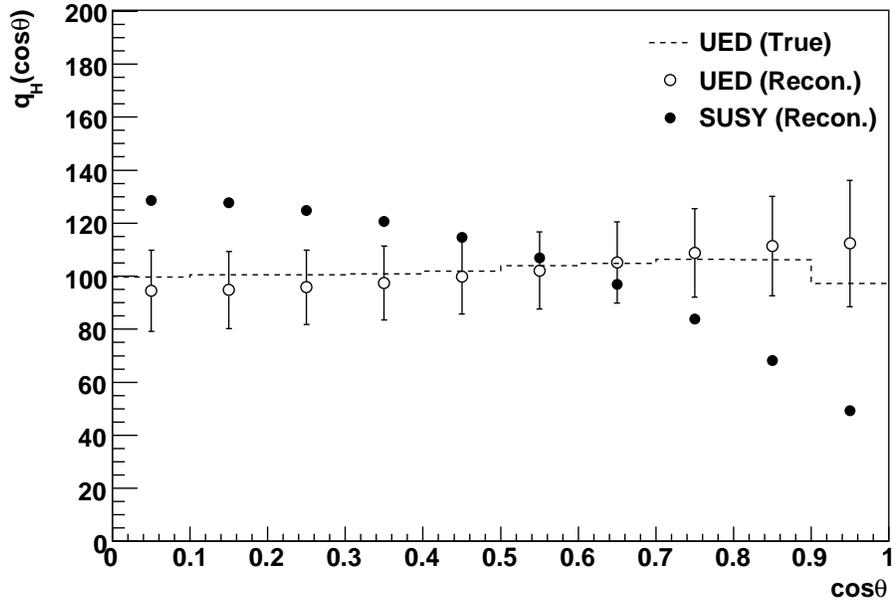}
    \label{f.cA.ued}
  }
  \caption{\small The reconstructed $q_H(\cos\theta_{\ma})$ distributions for events from the SUSY (filled circles) and UED models (open circles).}
  \label{f.cA}
\end{figure}

To begin, we will discuss our results from an attempt to reconstruct
the $\cos\theta_{\ma}$ distribution, which we divided into $M=10$
bins. To perform the reconstruction we assumed masses
$m_{\chi}=\hat{m}_{\chi}$ and $m_{\ma}=\hat{m}_{\ma}$, where
$\hat{m}_{\chi}=50 \text{ GeV}$ and $\hat{m}_{\ma} = 350\text{ GeV}$
are the true (pole) masses of $\chi$ and $\ma$, respectively, that
were used to generate the events. In Sec.~\ref{s.massErr} we will
discuss the effect of errors in these assumed masses.

In Fig.~\ref{f.cA.susy} and Fig.~\ref{f.cA.ued} we show the average
$\hat{q}_H(\cos\theta_{\ma})$ distribution for $H={\rm SUSY}$ and
$H={\rm UED}$, respectively. In each plot we show the distribution
obtained from events generated for the SUSY model (filled circles) and
UED model (open circles). The histogram in each plot shows the
$\langle N_S \rangle p(\cos\theta_{\ma}|H)$ distribution. Also shown
in each plot are error bars, where the error $\sigma^i$ in the $i^{\rm
  th}$ bin is given by the corresponding estimated variance:
\beq
\sigma^i = \sqrt{V_H^{ii}}
\eeq

It can be seen from these figures that events reconstructed with the true hypothesis have a $q_H$ distribution that is in agreement with $p(\cos\theta_{\ma}|H)$. This is as we expected, and provides an important validation of our method. However, we observe that there is a small discrepancy for events from the UED model. We expect that this is caused by our neglect of spin-correlations in the reconstruction, as discussed in Sec.~\ref{s.Implementation}. 

\begin{figure}[t]
  \centering
  \includegraphics[width=0.75\textwidth]{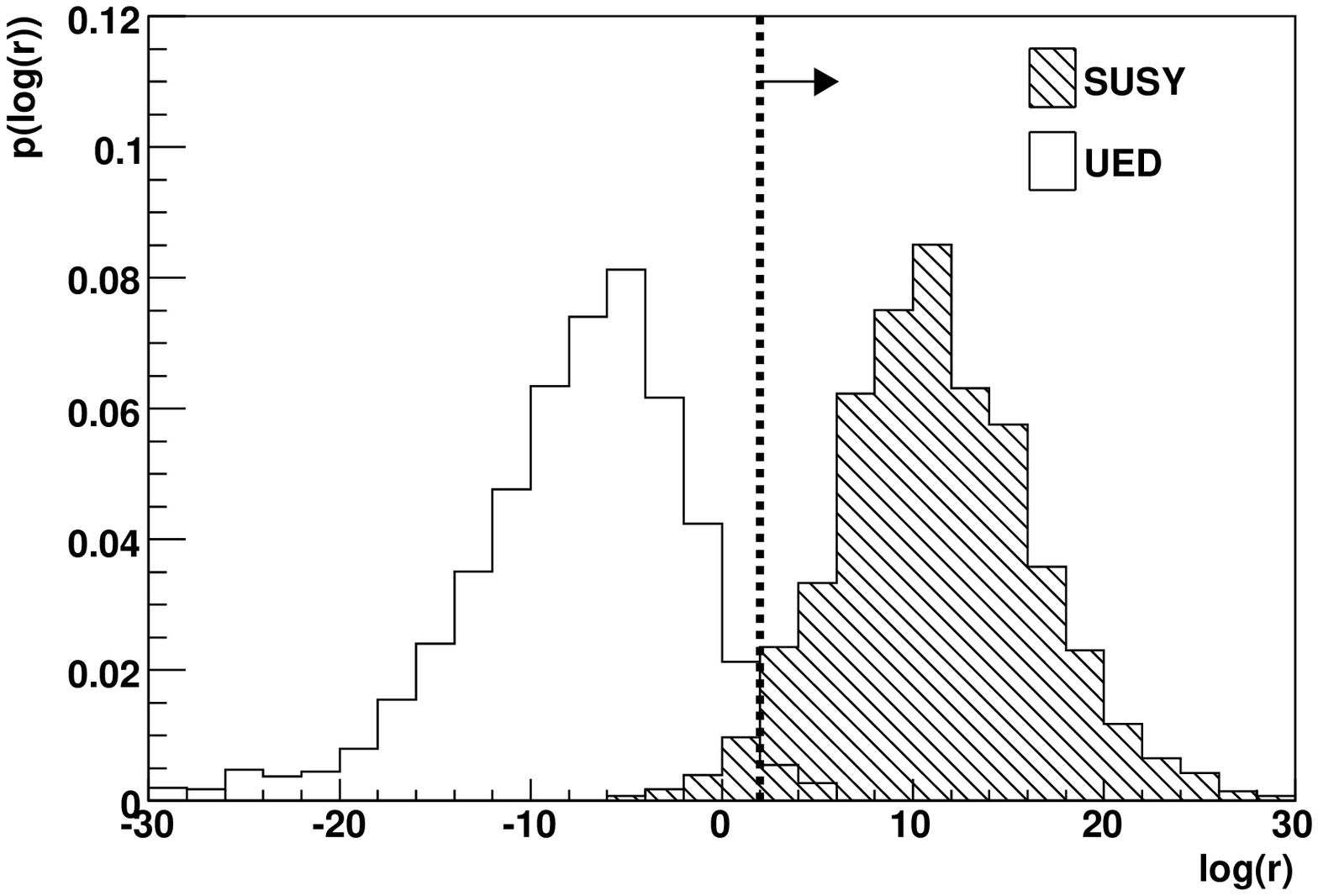}
  \caption{\small The $r$ distribution for events from the SUSY (hatched histogram) and UED (empty histogram) models, constructed using the $q_H(\cos\theta_{\ma})$ distribution. The dashed line and arrow indicate the acceptance region for the SUSY hypothesis, given in Eq.~\eqref{e.cA.acc}.}
  \label{f.cA.logr}
\end{figure}
Crucially, we observe that there is a significant discrepancy between events from the SUSY and UED models when reconstructed with the same hypothesis. This demonstrates that the method can discriminate between the two models, at least in principal. 
In Fig.~\ref{f.cA.logr} we show a plot of the $r$ distribution for events from SUSY (hatched histogram) and UED (empty histogram). The SUSY and UED distributions are well separated, thus indicating that $r$ is a potentially useful test statistic. To define our statistical test we choose our acceptance region, in which we accept the SUSY hypothesis instead of UED, as:
\beq\label{e.cA.acc}
\ln(r)>2
\eeq
From the distribution of Fig.~\ref{f.cA.logr} this test is expected to have a significance\footnote{The significance of a hypothesis test is defined as the probability of an error of the first kind, i.e. the probability of rejecting the hypothesis of SUSY, if SUSY is the true hypothesis~\cite{Cowan:1998ji}.}:
\beq\label{e.cA.sig}
\alpha = \int_{- \infty}^{2} p(\ln(r) | {\rm SUSY}) d\ln(r) = 3.3\%
\eeq
and power\footnote{The power of a hypothesis test to discriminate between the hypotheses is given by $(1-\beta)$, where $\beta$ is the probability of an error of the second kind, i.e. the probability of accepting the hypothesis of SUSY when SUSY is false~\cite{Cowan:1998ji}.}:
\beq\label{e.cA.power}
1-\beta = 1-\int_{2}^{\infty} p(\ln(r) | {\rm UED}) d\ln(r) = 98.3\%
\eeq

\begin{figure}[p]
  \centering
  \subfigure[$\cos\theta_{ll}$]{
    \includegraphics[width=0.85\textwidth]{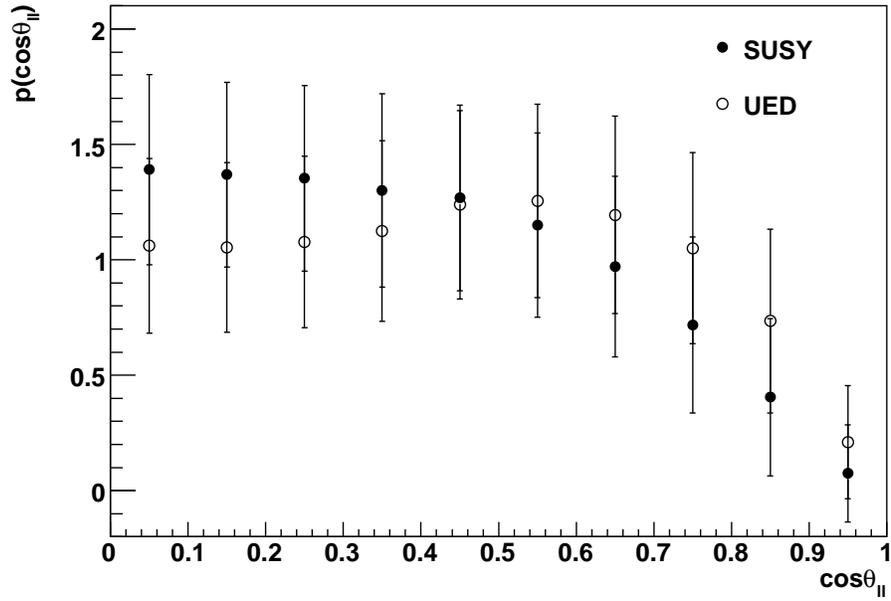}
    \label{f.cll}
  }
  \subfigure[$\log(r_{ll})$]{
    \includegraphics[width=0.85\textwidth]{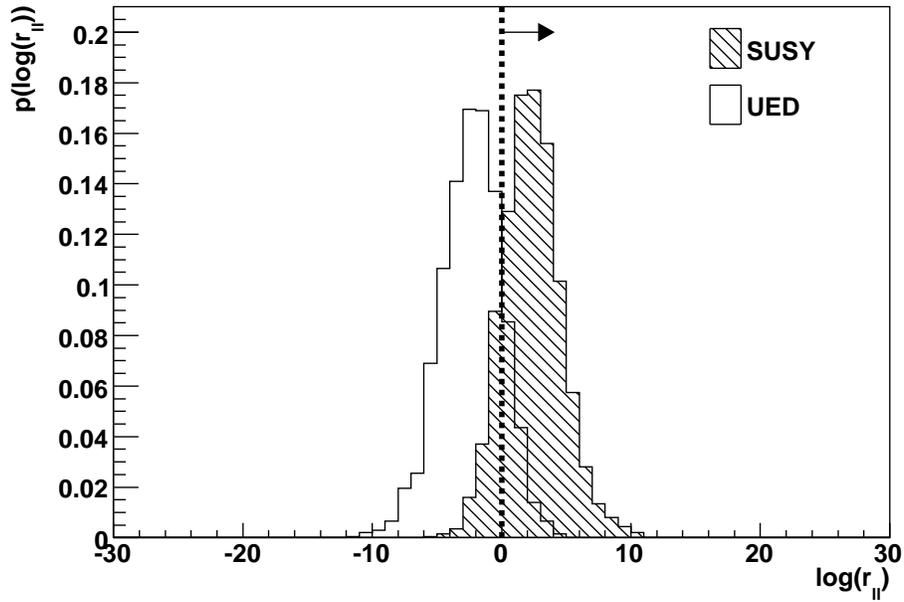}
    \label{f.cll.logr}
  }
  \caption{\small The $\cos\theta_{ll}$ and $r_{ll}$ distributions for events from the SUSY (filled circles, hatched histogram) and UED (open circles and empty histogram) models. The dashed line and arrow indicate the acceptance region for the SUSY hypothesis, given in Eq.~\eqref{e.cll.acc}.}
  \label{f.cll.all}
\end{figure}
To put this result into context we compared this with a test based upon the $\cos\theta_{ll}$ distribution. In Fig.~\ref{f.cll} we show the average $\cos\theta_{ll}$ distribution for our sample of 2000 pseudo-experiments, for both SUSY and UED models, after the average SM background contribution has been subtracted. 

We constructed a statistical test for this method as follows. We assumed, for each hypothesis $H$, that the number of events $n_i^H$ in the $i^{\rm th}$ bin, after subtraction of the averaged background, is a Gaussian random variable with a mean $\mu_i^H$ and variance $(\sigma_i^H)^2$ that were estimated from the set of pseudo-experiments. Using these Gaussian distributions we calculated for each pseudo-experiment the likelihood ratio 
\begin{align}
r_{ll} &= \frac{p(n_1, ... , n_{10} | {\rm SUSY})}{p(n_1, ..., n_{10}|{\rm UED})} \\
& = \prod_{i=1}^{10}\frac{\sigma_i^{\rm UED}}{\sigma_i^{\rm SUSY}}\exp\bigg(-\frac{(\mu_i^{\rm SUSY}-n_i)^2}{2(\sigma_i^{\rm SUSY})^2}+\frac{(\mu_i^{\rm UED}-n_i)^2}{2(\sigma_i^{\rm UED})^2}\bigg)
\end{align}
The distribution of $r_{ll}$ is shown for both SUSY and UED models in Fig.~\ref{f.cll.logr}. The SUSY and UED distributions are separated, however there is a significant overlap between the two. If we define the acceptance region as:
\beq\label{e.cll.acc}
\ln(r_{ll})>0
\eeq
we find a statistical test with acceptance $\alpha=14.8\%$ and power $1-\beta=84.9\%$. Comparing this with Eqns.~\eqref{e.cA.sig} and \eqref{e.cA.power} we find that our method reduces the probability of error by a factor of $\sim 5$. 

\begin{figure}[p]
  \centering
  \subfigure[$H=\text{SUSY}$]{
    \includegraphics[width=0.85\textwidth]{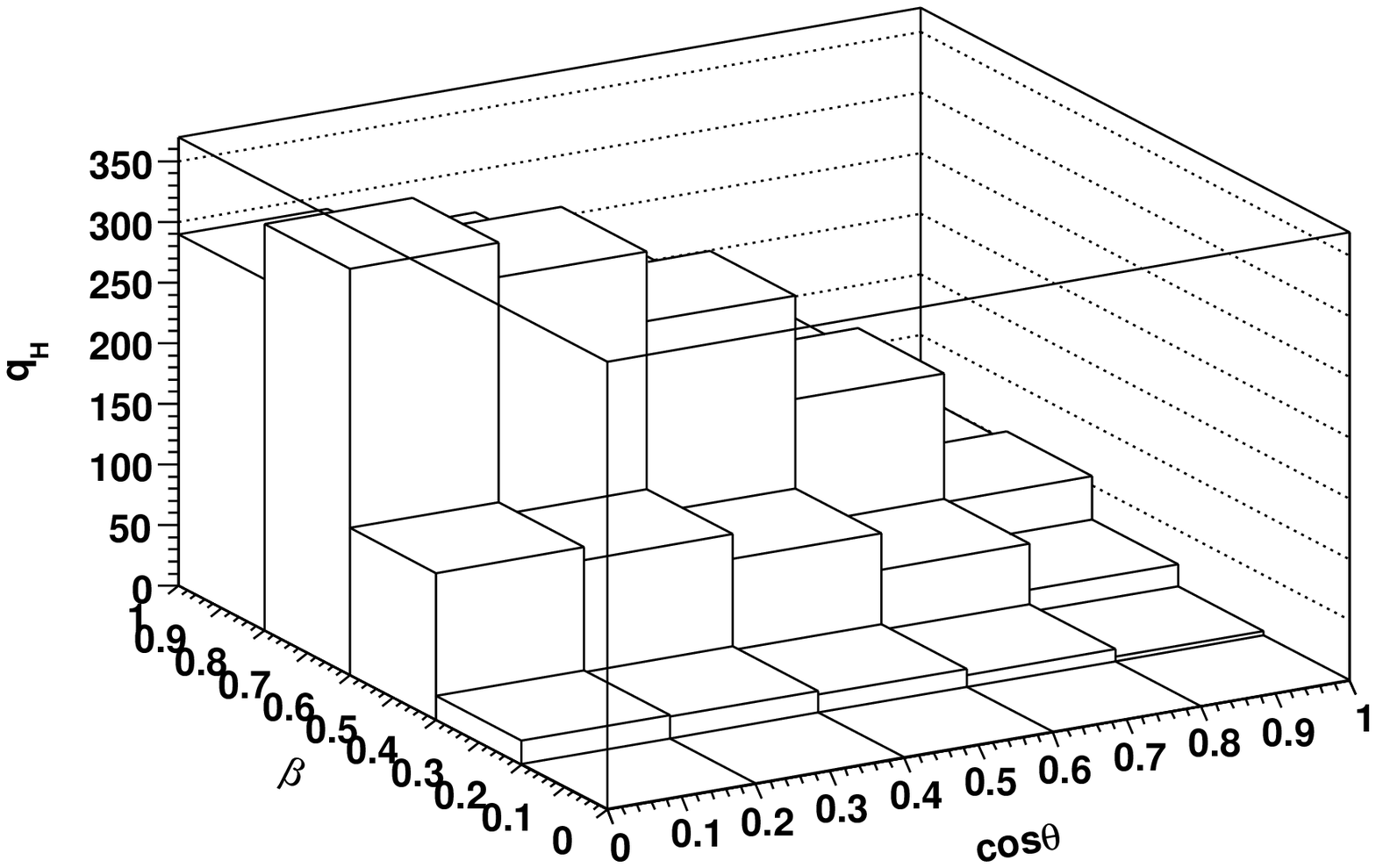}
    \label{f.susy.cYbY.susy}
  }
  \subfigure[$H=\text{UED}$]{
    \includegraphics[width=0.85\textwidth]{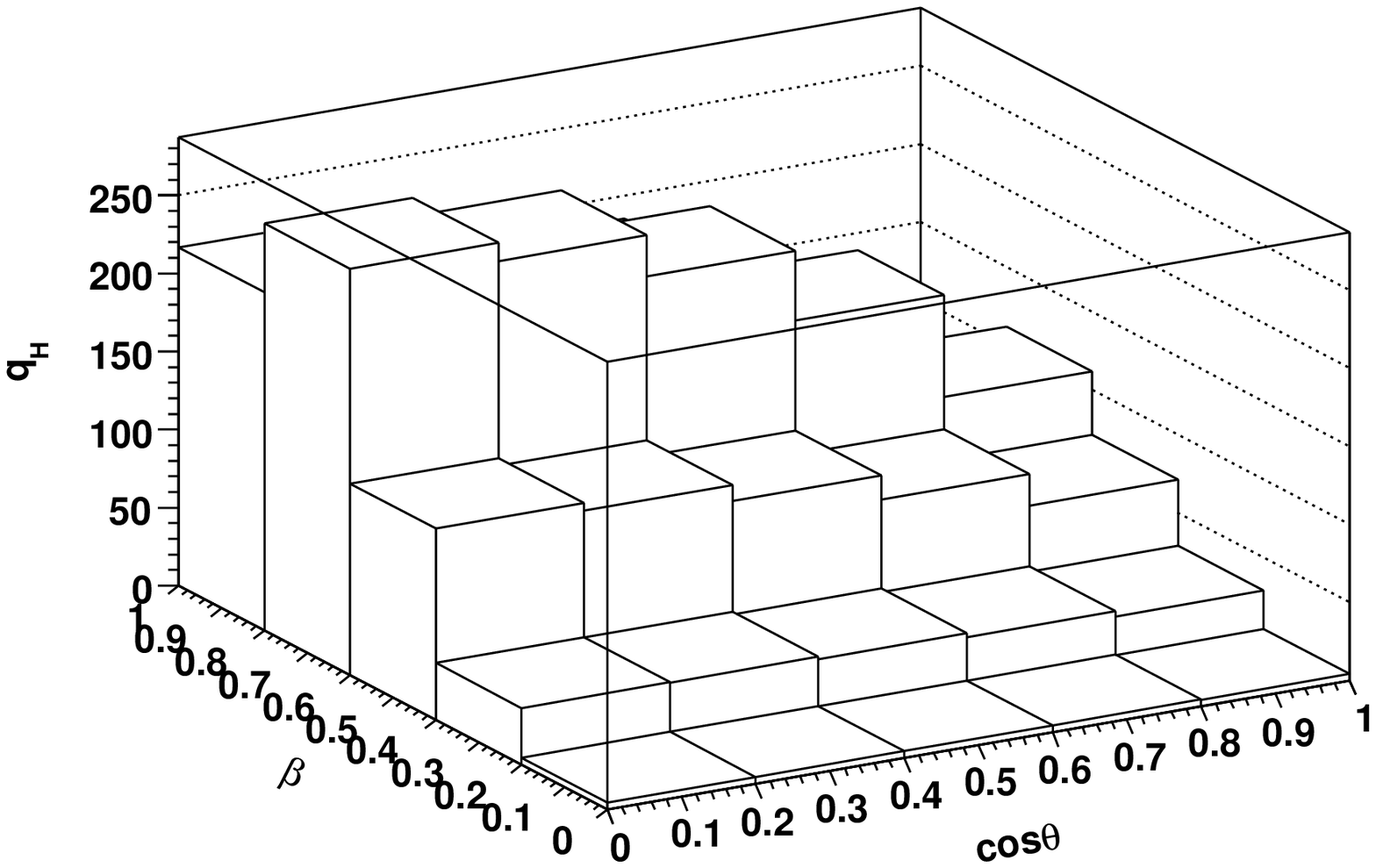}
    \label{f.susy.cYbY.ued}
  }
  \caption{\small The average joint $q_H(\cos\theta_{\ma}, \beta_{\ma})$ distribution for SUSY events, after background subtraction.}
  \label{f.susy.cYbY}
\end{figure}
\begin{figure}[p]
  \centering
  \subfigure[$H=\text{SUSY}$]{
    \includegraphics[width=0.85\textwidth]{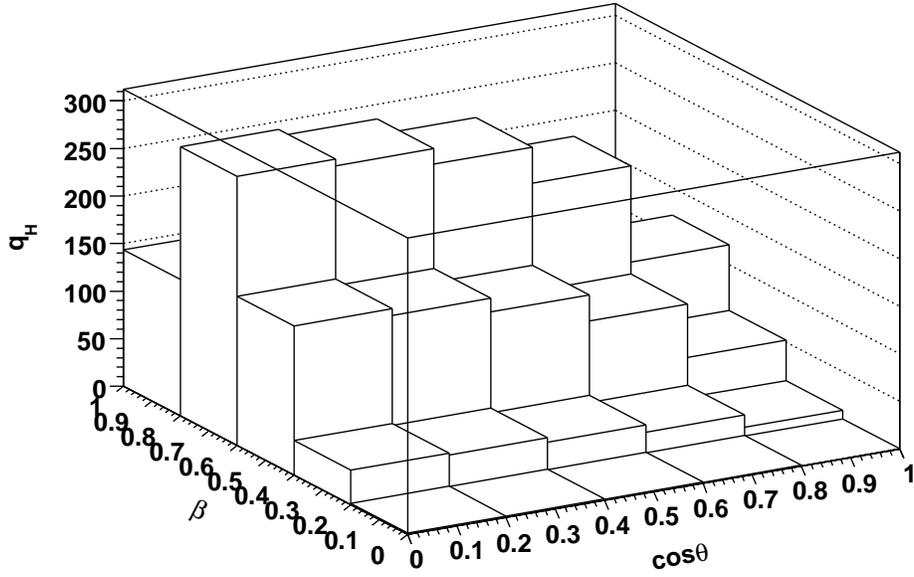}
    \label{f.ued.cYbY.susy}
  }
  \subfigure[$H=\text{UED}$]{
    \includegraphics[width=0.85\textwidth]{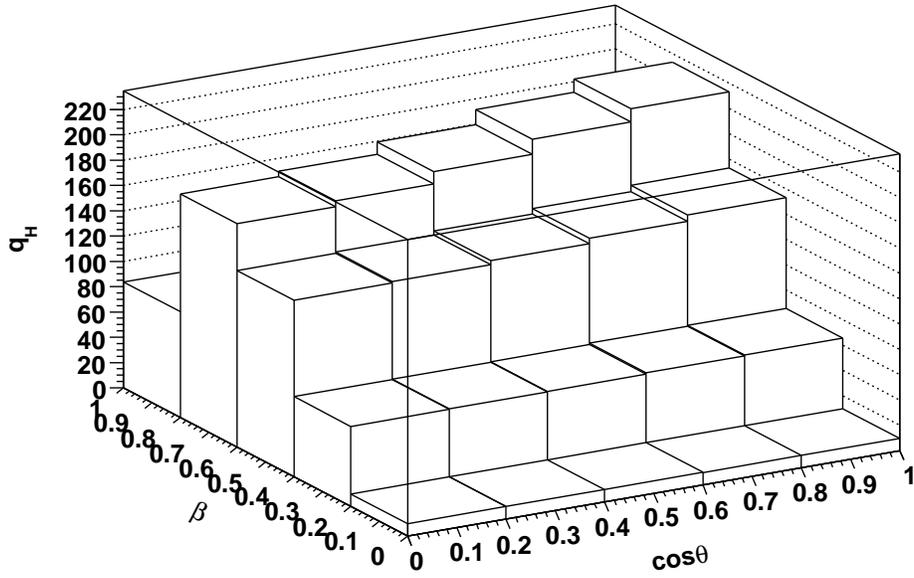}
    \label{f.ued.cYbY.ued}
  }
  \caption{\small The average joint $q_H(\cos\theta_{\ma}, \beta_{\ma})$ distribution for UED events, after background subtraction.}
  \label{f.ued.cYbY}
\end{figure}
\begin{figure}[t]
  \centering
  \includegraphics[width=0.75\textwidth]{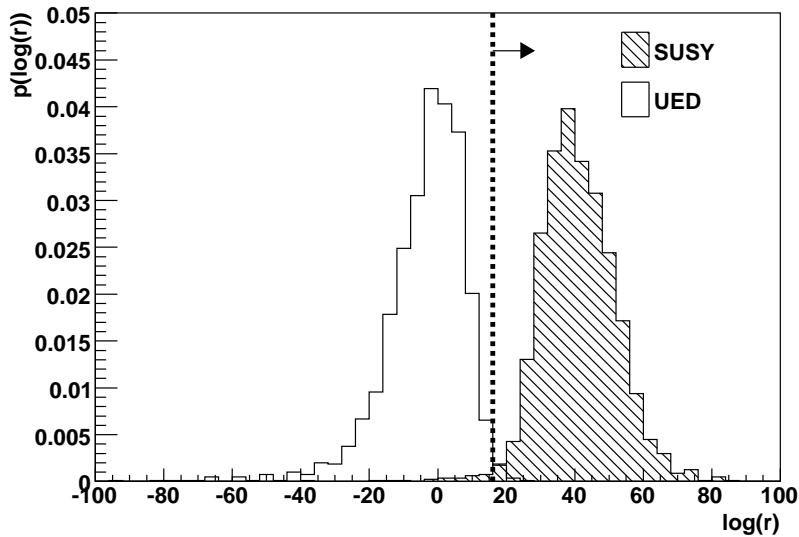}
  \caption{\small The $r$ distribution for SUSY (hatched) and UED (empty) events, constructed from the joint $q_H(\cos\theta_{\ma}, \beta_{\ma})$ distribution. The dashed line and arrow indicate the acceptance region for the SUSY hypothesis, given in Eq.~\eqref{e.cAbA.acc}.}
  \label{f.cYbY.logR}
\end{figure}

Following this initial test we attempted to improve our model discrimination by including additional kinematic variables. In Fig.~\ref{f.susy.cYbY} we show the joint distribution $q_H(\cos\theta(\ma), \beta_{\ma})$ for the SUSY events, reconstructed with both hypotheses after the average background has been subtracted. The distribution for the UED events is shown in Fig.~\ref{f.ued.cYbY}. One interesting feature of these distributions is that the $\beta_{\ma}$ dependence is insensitive to the hypothesis $H$ used in the reconstruction. As can be observed in Fig.~\ref{f.susy.cYbY} the $\beta_{\ma}$ distribution for the SUSY events is peaked closer to 1 than in the UED events, as one would expect given the $\beta_{\ma}$ dependence of the matrix elements. Following the same procedure as before we constructed the probability distributions $f(\vec{q}_H|H)$, and determined the likelihood ratio $r$ for each experiment. This $r$ distribution is given in Fig.~\ref{f.cYbY.logR} for both SUSY and UED events.

Comparing Fig.~\ref{f.cYbY.logR} and Fig.~\ref{f.cA.logr} one can see that by combining both $\cos\theta_{\ma}$ and $\beta_{\ma}$ it is possible to enhance the separation of the SUSY and UED events, compared to reconstructing $\cos\theta_{\ma}$ alone. Defining the acceptance region for the SUSY hypothesis as:
\beq\label{e.cAbA.acc}
\ln(r)>16.
\eeq
we obtain a statistical test with significance $\alpha = 1.2\%$ and power $1-\beta = 99.1\%$. This test improves over using $\cos\theta_{\ma}$ alone by a factor of $\sim2$, and is over 10 times less likely to result in error than the $\cos\theta_{ll}$ method. 

Attempts to further improve the hypothesis test by including the kinematic variables $\phi$, $\theta_{\mu}$ and $\bar{\theta}_{\mu}$ were unsuccessful. 
Although our method could reconstruct some features of these distributions they were not statistically significant once the background contribution was taken into account.

\subsubsection{The effect of mass uncertainties} \label{s.massErr}
Errors in the measurement of the masses $m_{\chi}$ and $m_{\ma}$ will affect the efficacy of our proposed statistical tests. To study the effect of such a systematic error we varied the true mass $\hat{m}_{\ma}$, used to generate the events, whilst keeping our statistical test fixed. (That is, the masses assumed for the reconstruction were fixed at $m_{\chi}=50\text{ GeV}$ and $m_{\ma}=350\text{ GeV}$, and we used the same probability densities $f(q_H|H)$ and acceptance region defined in the previous section.) 
Assuming an $\mathcal{O}(10\%)$ uncertainty in the mass $m_{\ma}$, we performed another 2000 pseudo-experiments for the cases $\hat{m}_{\ma}=1.1m_{\ma}$ and $\hat{m}_{\ma}=0.9m_{\ma}$. For each case, the significance and power of each statistical test is given in Table~\ref{t.Results}.

We observe that the tests based upon the $q_H(\cos\theta_{\ma})$ and $\cos\theta_{ll}$ distributions are quite insensitive to the 10\% mass uncertainty, with variations in $\alpha$ and $\beta$ at the level of 10---50\%. Thus these tests, at least, are fairly robust against uncertainties in the mass measurement. However, the test based upon $q_H(\cos\theta_{\ma}, \beta_{\ma})$ exhibits a much larger variation with $\beta=0.9^{+ 3.9}_{-0.6}\%$ and $\alpha=1.2^{+1.6}_{-0.7}\%$. This large variation appears to be caused by the sensivity of the $q_H(\beta_{\ma})$ distribution to the true and assumed masses. Hence, a 10\% mass uncertainty reduces the significance and power of this test, causing it to be less effective than the test based upon $q_H(\cos\theta_{\ma})$.

\begin{table}[tb]
  \begin{center}
  \begin{tabular}{|c|c|c|c|c|c|c|c|}
    \hline 
    \multirow{2}{*}{Variable} & \multirow{2}{*}{Acceptance Region} &  \multicolumn{2}{|c|}{$\hat{m}_{\ma} = m_{\ma}$} & \multicolumn{2}{|c|}{$\hat{m}_{\ma} = 0.9m_{\ma}$} & \multicolumn{2}{|c|}{$\hat{m}_{\ma} = 1.1m_{\ma}$}  \\
    \cline{3-8}
    & & $\alpha$ & $1-\beta$ & $\alpha$ & $1-\beta$ & $\alpha$ & $1-\beta$ \\
    \hline                                                                                            $q_H(\cos\theta_{\ma})$ & $\ln(r)>2$ & 3.3\% & 98.3\% & 3.3\% & 96.1\% & 4.7\% & 98.5\%  \\
    $q_H(\cos\theta_{\ma}$, $\beta_{\ma})$  & $\ln(r)>16$ & 1.2\% & 99.1\% & 2.8\% & 99.7\%& 0.5\%     &  95.2\%    \\
    $\cos\theta_{ll}$ & $\ln(r_{ll})>0$ &  14.8\% & 84.9\% & 12.3\% & 84.6\% & 15.3\% & 82.7\%  \\
    \hline
  \end{tabular}
  \end{center}
  \caption{\small The significance $\alpha$ and power $1-\beta$ of several statistical tests used to discriminate between SUSY and UED models, for various true masses $\hat{m}_{\ma}$.}
  \label{t.Results}
\end{table}

\section{Conclusions} \label{s.Conclusion}
In this paper we have proposed a new method for reconstructing the probability distributions of kinematic variables that are sensitive to the spin of new particle states. This method relies on performing a probabilistic reconstruction of each event, combining information from the measured particle masses and the matrix element for a hypothesised process. As a result, one is able to study kinematic observables that depend on unknown momenta, such as that carried by two potential DM candidates. By combining the information from a sample of events we have shown how one can attempt to reconstruct this probability distribution and then test the assumed hypothesis.

We have performed a preliminary test of the method on a candidate process: the discrimination of SUSY and UED models in slepton-pair production. This test was performed at parton level, without hadronisation or detector simulation, and took into account the effects of event selection and the dominant SM backgrounds. We have shown that with 100$\text{ fb}^{-1}$ of luminosity one can discriminate between the SUSY and UED models, with a probability of error of $\mathcal{O}(3\%)$. This result has been shown to be robust against a 10\% uncertainty in the mass measurement. Our complete results are summarised in Table~\ref{t.Results} and show that, in principle, one can improve the test by combining several independent variables. However, the improvement gained in practice will depend upon the uncertainty in the masses.

We compared this result with that expected for an alternative method, which uses the variable $\cos\theta_{ll}$. We find, for the process considered, that this can perform a discrimination with a probability of error of $\mathcal{O}(15\%)$. Thus, potentially, our method offers an improved way of discriminating between models and measuring the spin.

Despite this encouraging result, many further questions remain and deserve detailed study. The true efficacy of our method can only be understood through a real world application, together with more detailed simulation. In particular, it will be crucial to study the effect of systematic uncertainties in the background estimation and mass measurement, combined with the effects of detector resolution and acceptance. Furthermore, it would be of interest to study how this method performs in a wider range of processes. Possible processes of interest are those that involve long decay chains, or those involving diagrams with both s- and t-channel particle exchange.

\section*{Acknowledgements}
DH would like to thank Graham G. Ross for several enlightening discussions and comments on the manuscript. Further thanks also go to Alan J. Barr, for useful insights on experimental matters related to event selection and simulation. This research project has been supported by a Marie Curie Early Initial Training Network Fellowship of the European Community's Seventh Framework Programme under contract number (PITN-GA-2008-237920-UNILHC).

\bibliographystyle{elsart-num}
\bibliography{eRbib}

\begin{thebibliography}{10}
\expandafter\ifx\csname url\endcsname\relax
  \def\url#1{\texttt{#1}}\fi
\expandafter\ifx\csname urlprefix\endcsname\relax\def\urlprefix{URL }\fi

\bibitem{Athanasiou:2006ef}
C.~Athanasiou, C.~G. Lester, J.~M. Smillie, B.~R. Webber, {Distinguishing spins
  in decay chains at the Large Hadron Collider}, JHEP 08 (2006) 055.

\bibitem{Barr:2004ze}
A.~J. Barr, {Using lepton charge asymmetry to investigate the spin of
  supersymmetric particles at the LHC}, Phys. Lett. B596 (2004) 205--212.

\bibitem{Wang:2006hk}
L.-T. Wang, I.~Yavin, {Spin Measurements in Cascade Decays at the LHC}, JHEP 04
  (2007) 032.

\bibitem{Burns:2008cp}
M.~Burns, K.~Kong, K.~T. Matchev, M.~Park, {A General Method for
  Model-Independent Measurements of Particle Spins, Couplings and Mixing Angles
  in Cascade Decays with Missing Energy at Hadron Colliders}, JHEP 10 (2008)
  081.

\bibitem{Kramer:2009kp}
M.~Kramer, E.~Popenda, M.~Spira, P.~M. Zerwas, {Gluino Polarization at the
  LHC}, Phys. Rev. D80 (2009) 055002.

\bibitem{Barr:2005dz}
A.~J. Barr, {Measuring slepton spin at the LHC}, JHEP 02 (2006) 042.

\bibitem{Boudjema:2009fz}
F.~Boudjema, R.~K. Singh, {A model independent spin analysis of fundamental
  particles using azimuthal asymmetries}, JHEP 07 (2009) 028.

\bibitem{Choi:2006mr}
S.~Y. Choi, K.~Hagiwara, H.~U. Martyn, K.~Mawatari, P.~M. Zerwas, {Spin
  analysis of supersymmetric particles}, Eur. Phys. J. C51 (2007) 753--774.

\bibitem{Buckley:2008eb}
M.~R. Buckley, S.~Y. Choi, K.~Mawatari, H.~Murayama, {Determining Spin through
  Quantum Azimuthal-Angle Correlations}, Phys. Lett. B672 (2009) 275--279.

\bibitem{Barr:2010zj}
A.~J. Barr, C.~G. Lester, {A Review of the Mass Measurement Techniques proposed
  for the Large Hadron Collider}.

\bibitem{Lester:1999tx}
C.~G. Lester, D.~J. Summers, {Measuring masses of semiinvisibly decaying
  particles pair produced at hadron colliders}, Phys. Lett. B463 (1999)
  99--103.

\bibitem{Barr:2003rg}
A.~Barr, C.~Lester, P.~Stephens, {m(T2) : The Truth behind the glamour}, J.
  Phys. G29 (2003) 2343--2363.

\bibitem{Aad:2009wy}
G.~Aad, et~al., {Expected Performance of the ATLAS Experiment - Detector,
  Trigger and Physics}.

\bibitem{Acosta:922757}
D.~Acosta, M.~Della~Negra, L.~Foà, A.~Hervé, A.~Petrilli, CMS physics:
  Technical Design Report, Technical Design Report CMS, CERN, Geneva, 2006.

\bibitem{Alwall:2007st}
J.~Alwall, et~al., {MadGraph/MadEvent v4: The New Web Generation}, JHEP 09
  (2007) 028.

\bibitem{Meade:2007js}
P.~Meade, M.~Reece, {BRIDGE: Branching ratio inquiry / decay generated events}.

\bibitem{Martin:2009iq}
A.~D. Martin, W.~J. Stirling, R.~S. Thorne, G.~Watt, {Parton distributions for
  the LHC}, Eur. Phys. J. C63 (2009) 189--285.

\bibitem{Cowan:1998ji}
G.~Cowan, Statistical data analysis, Oxford University Press, Oxford, UK, 1998.

\end{thebibliography}

\end{document}